\documentclass[12pt]{article}

\usepackage{%color,
graphicx,amsmath,amssymb,bm}

\newtheorem{prop}{Prop}[section]
\newtheorem{lemma}[prop]{Lemma}
\newtheorem{definition}[prop]{Definition}
\newtheorem{theorem}[prop]{Theorem}

\newcommand{\qed}{~ \rule{0.8ex}{1.6ex}\smallskip}

\title{Universal nature of replica symmetry breaking in quantum systems with Gaussian disorder}
\author{C. Itoi \\ 
Department of Physics, GS and CST, Nihon University, \\
Kanda-Surugadai, Chiyoda, Tokyo 101-8308, Japan} 
\begin{document}
\maketitle
\begin{abstract}{
We study quantum spin systems with quenched Gaussian disorder. 
We prove that the variance of all physical quantities in a certain class 
vanishes in the infinite volume limit.  
We study also replica symmetry breaking phenomena, where the variance  of an overlap operator in the other class
does not vanish
in the replica symmetric Gibbs state. 
On the other hand,  it vanishes in a spontaneous replica symmetry breaking Gibbs state defined by applying 
an infinitesimal  replica symmetry breaking field.
We prove also that the finite variance of the overlap operator in the replica symmetric Gibbs state 
implies the existence of a spontaneous replica symmetry breaking.  
% \PACS{PACS code1 \and PACS code2 \and more}
% \subclass{MSC code1 \and MSC code2 \and more
}
\end{abstract}

%\date{Received: date / Accepted: date}
% The correct dates will be entered by the editor

%
\section{Introduction}
Coupling constants  in a system with quenched disorder are given by  i.i.d. random variables.
We can regard a given disordered sample as a system obtained by a random sampling of these variables.
All physical quantities in such systems are functions of these random variables.
In statistical physics, a physical quantity is said to be self-averaging, 
if its observed value is equal to its expectation value in a disordered sample. 
In other words, self-averaging quantities obey the law of large numbers. 
In several disordered spin systems, the free energy density 
is known to be self-averaging.
To be specific,  first we consider disordered Ising systems. 
Let $\Lambda_L:= [1,L]^d \cap {\mathbb Z}^d
$ be a lattice whose volume is $|\Lambda_L|=L^d$.
We define a collection $C_L$ of interaction ranges $X \subset \Lambda_L$
%$S =(S_x)_{x \in \Lambda_L}$ be a spin configuration of the Ising spin $S_x = \pm 1$ at a site $x \in \Lambda_L$.
For an arbitrary  $X \in C_L$, we denote 
$$S_X = \prod_{x \in X} S_x.$$     
For coupling constants $J_1 , J_0, h \in {\mathbb R}$. we define  Hamiltonian by
$$
H(S,g) :=-\sum_{X \in C_L} ( J_1 g_X+J_0)S_X -h \sum_{x\in \Lambda_L} S_x
$$ 
as a function of the spin configuration and
 i.i.d. standard Gaussian random variables $g=(g_X)_{ X \in C_L}$.
If the Hamiltonian is invariant under the following transformation
$$
H(-S,g)=H(S,g).
$$ 
 for $h=0$, this symmetry is  called ${\mathbb Z}_2$ symmetry.
In such systems, the magnetization density 
$$m_L := \frac{1}{|\Lambda_L|} \sum_{x\in \Lambda_L} S_x,$$
of spin variable $S_x = \pm 1$ is an order parameter which is 
a random variable satisfying the Gibbs distribution depending on
the quenched disorder $g=(g_X)_{ X \in C_L}$.
Griffiths' theorem  \cite{Gff} for the system without disorder $J_1=0$ shows that the long range order
$
\lim_{L \uparrow \infty} \lim_{h \downarrow 0}  \langle \delta m_L ^2 \rangle \neq 0
$
should imply the spontaneous symmetry breaking 
$
 \lim_{h \downarrow 0}  \lim_{L \uparrow \infty} \langle m_L \rangle \neq 0,
$
where  $\langle \cdots \rangle$ is expectation in the Gibbs state and a deviation is defined by
$\delta m_L := m_L -\langle m_L \rangle.$ 
This is believed for the spontaneous ${\mathbb Z}_2$ symmetry breaking also in disordered systems.
Define another deviation 
$\Delta m_L := m_L -{\mathbb E} \langle m_L \rangle,$ 
and consider the following variance
$${\mathbb E}\langle \Delta m_L^2 \rangle :={ \mathbb E}\langle (m_L -{\mathbb E} \langle m_L  \rangle )^2 \rangle ,$$
 where ${\mathbb E}$ is the sample expectation
with respect to the quenched disorder. 
By the Chebyshev inequality,  the variance gives
an upper bound of the probability that a difference between
the observed value $m_L$ and its expectation value $ {\mathbb E} \langle m_L\rangle$ becomes 
larger than a positive number $c$, 
$$
P(| m_L   -{\mathbb E} \langle m_L\rangle| >c  ) \leq \frac{{\mathbb E}\langle \Delta m_L^2\rangle}{c^2 }.
$$  
If the  variance 
 %obeys the law of large numbers.
vanishes in the infinite volume limit 
$$
\lim_{L \rightarrow \infty}{\mathbb E}\langle \Delta m_L^2\rangle = 0,
$$  
the observed value of the magnetization density $m_L$ differs  from its expectation value rarely.
In this case, $m_L$ is self-averaging. 

Next, we consider 
replica symmetry breaking phenomena 
which apparently violate self-averaging 
of the overlap between two replicated quantities in a replica symmetric expectation.   
%First, we explain the replica symmetry and its breaking in 
%the Sherrington-Kirkpatrick model, as an example of disordered classical Ising systems on a lattice $\Lambda_L$. 
%the quenched disorder $J_X$ is written in terms of standard Gaussian random variable $g_X$, 
Let  $S^\alpha (\alpha=1, \cdots, n)$ be $n$ replicated copies of a spin configuration, and we consider the following Hamiltonian 
$$
H(S^1, \cdots, S^n, g):= \sum_{a=1}^n H(S^\alpha, g),
$$
where replicated spin configurations share the same quenched disorder $g$.
This Hamiltonian is invariant under an arbitrary permutation $\sigma \in S_n$.
$$ H(S^1, \cdots, S^n g)=H(S^{\sigma1}, \cdots, S^{\sigma n} , g)$$
 
This permutation symmetry is the replica symmetry.
The spin overlap $R_{1,2}$ between two replicated spin configurations is defined by   
$$
R_{1,2}:=\frac{1}{|C|}\sum_{X \in C_L} S_X^1 S_X^2.
$$
The covariance of the Hamiltonian is written in terms of the overlap
$$
{\mathbb E} H(S^1,g) H(S^2,g) - {\mathbb E} H(S^1,g) {\mathbb E}H(S^2, g)=  |C| J_1^2R_{1,2}.
$$
%where  and $C_L$ is a collection of interaction ranges.  
When the replica symmetry breaking occurs, broadening of the overlap distribution with a finite variance is observed.
This phenomenon is  well-known in several disordered systems, such as the Sherrington-Kirkpatrick model \cite{Pr,T2,T}. 
In the present paper, we define replica symmetry breaking by the finite  variance  calculated in the replica symmetric expectation
in the infinite volume limit
$$
\lim_{L \rightarrow \infty} {\mathbb E}  \langle \Delta R_{1,2} ^2 \rangle
 =\lim_{L \rightarrow \infty} [
 {\mathbb E}  \langle R_{1,2} ^2 \rangle - ({\mathbb E }  \langle R_{1,2} \rangle  )^2 ]\neq 0,
$$
where
$\Delta R_{1,2} := R_{1,2} -{\mathbb E }  \langle R_{1,2} \rangle  $.  
This definition of the replica symmetry breaking is given by  Chatterjee \cite{C2}. 
Although the replica symmetry breaking is believed to be a spontaneous symmetry breaking, 
there are not many studies of the replica symmetry breaking regarded as
spontaneous  symmetry breaking \cite{G2,G3,M2}. There are lots of important studies for 
spontaneous symmetry breaking for spin systems without disorder.   
Griffiths' theorem %for  is one of them.This theorem 
shows that  the spontaneous symmetry breaking in ferromagnetic systems can be detected
by the  finite variance of the order parameter calculated in the symmetric Gibbs state \cite{Gff}. 
Namely, the long range order implies a finite spontaneous magnetization.
%In the present paper, we %study spontaneous replica symmetry breaking . We 
%clarify the relation  between the finite variance of overlap %$\delta R_{1,2}$ and $\Delta R_{1,2}$  
%and  the spontaneous replica symmetry breaking in disordered spin systems. 
In a certain sample with a quenched disorder $g$, we are interested in
 %expect that the spontaneous  replica  symmetry breaking can be detected by 
another variance
$$
\lim_{L \rightarrow \infty}  \langle \delta R_{1,2} ^2 \rangle,
$$
calculated in the replica symmetric Gibbs state,
where 
$\delta R_{1,2} :=R_{1,2}  -  \langle R_{1,2} \rangle.$ 
If this variance does not vanish, this phenomenon can be regarded  as a long range order.  
The following  finite variance in a sample expectation
$$
\lim_{L \rightarrow \infty}
{\mathbb E}  \langle \delta R_{1,2} ^2 \rangle
=\lim_{L \rightarrow \infty} [ 
{ \mathbb E }\langle R_{1,2} ^2 \rangle - { \mathbb E } \langle R_{1,2} \rangle^2]\neq 0,
$$ 
implies that  the long range order  is not a rare event. In this case, one can expect an instability of symmetric Gibbs state because of the strong fluctuation and  also spontaneous replica symmetry breaking. 
%rather than $\lim_{L \rightarrow \infty} {\mathbb E}  \langle \Delta R_{1,2} ^2 \rangle$,
%Here,  we  calculate the variance by replica symmetric Gibbs state.
 %The finiteness $\lim_{L \rightarrow \infty}{\mathbb E}  \langle \delta R_{1,2} ^2 \rangle>0$ suggests that samples with unstable replica symmetric Gibbs states are not rare.  %$\lim_{L \rightarrow \infty}{\mathbb E}  \langle \Delta R_{1,2} ^2\rangle$
%The spontaneous replica symmetry breaking apparently violates 
%the law of large number, since the variance of the overlap $R_{1,2}$ becomes  finite in the replica symmetric calculation. 
%Actually however, the replica symmetry breaking Gibbs state realizes instead of the 
%unstable replica symmetric one and the variance should vanish there. 

Next, we discuss a possibility of replica symmetry breaking without long range order.
Since
$$
\lim_{L \rightarrow \infty}{\mathbb E}  \langle \Delta R_{1,2} ^2 \rangle=\lim_{L \rightarrow \infty}
{\mathbb E}  \langle \delta R_{1,2} ^2 \rangle+\lim_{L \rightarrow \infty}{\mathbb E}  \langle \Delta R_{1,2} \rangle^2,
$$  
the replica symmetry breaking occurs, if  $\lim_{L \rightarrow \infty}{\mathbb E}  \langle \Delta R_{1,2} \rangle^2> 0$ 
even in the case $\lim_{L \rightarrow \infty}{\mathbb E}  \langle \delta R_{1,2} ^2 \rangle=0$.  
At least in principle, replica symmetry breaking may occur
even though long range order is a rare event.  
%Although almost all samples have no long range order, There is a possibility that
%the overlap $R_{1,2}=\langle R_{1,2}\rangle$ of an arbitrary  typical sample 
%differs from the sample expectation ${\mathbb E} \langle R_{1,2}\rangle$.   
For the disordered Ising systems, however, this possibility is ruled out by 
\begin{equation}
2{\mathbb E}  \langle \Delta R_{1,2} ^2 \rangle
=3{\mathbb E}  \langle \delta R_{1,2} ^2 \rangle
=6{\mathbb E}  \langle \Delta R_{1,2}  \rangle^2
\label{GG}
\end{equation}
as follows from the Aizenman-Contucci  \cite{AC,CG} and the Ghirlanda-Guerra identities \cite{C2,C1,CG2,CG3,GG}.  Therefore,  the replica symmetry breaking cannot occur without long range order in the  disordered Ising systems. 
%There are several arguments that the replica symmetry breaking is a spontaneous symmetry breaking phenomenon \cite{G2,G3,M2}.
To extend this argument to quantum spin systems with quenched disorder, 
however, the quantum mechanically extended Ghirlanda-Guerra identities do not yield the simple identities (\ref{GG}) because of 
the non-commutativity of the spin operators \cite{I}.  
In the present paper, we prove that the finite variance of the overlap in the replica symmetric calculation implies  spontaneous replica symmetry breaking.
Namely, we show that  only spontaneous symmetry breaking Gibbs state causes replica symmetry breaking.
%The broken symmetry can be observed  even in the replica symmetric Gibbs sates. 
%In this paper, we study self-averaging and its breaking for several quantities in quantum spin models 
%with quenched disorder.  %One of the most important  motivations to study disordered quantum systems, 
 
\section{Definitions}

Let $\Lambda_L$ be a $d$ dimensional cubic  lattice with a linear size $L$.
A spin operator $S_x^i (i = 1,2,3)$ at a site $x \in \Lambda_L$ acting on the Hilbert space ${\cal H}:=\bigotimes_{x \in \Lambda_L} {\cal H}_x$
is defined by a tensor product of the $2S+1$ dimensional self-adjoint matrix $S^i$ acting on ${\cal H}_x \simeq {\mathbb C}^{2S+1}$  and 
$2S+1$ dimensional identity matrices. 
 Spin operators $(S_x^i )_{x \in \Lambda_L, i = 1,2,3}$ satisfy the  following commutation relation 
$$
[S_x^i, S_y^j] = i \delta_{x,y}\sum_{k=1}^3\epsilon_{i,j,k}  S_x^k.
$$
The magnitude of each spin operator $(S_x^1,S_x^2,S_x^3)$ is
fixed by
$$
\sum_{i=1}^3 (S_x ^i)^2= S(S+1) \vec 1,
$$
where $\vec 1$ is  the identity  operator on the Hilbert space ${\cal H}$.
We denote a product of the spin operators
$$
S_X ^i = \prod_{x \in X} S_x^i.
$$ 
To define models, let $A:=\{a \in {\mathbb Z} | 1 \leq a \leq l \}$ be a finite set and
$(C_L^a)_{a \in A}$ be a collection of interaction ranges, where each $X \in C_L^a$ is a subset $X \subset \Lambda_L$ 
and $|X| = n_a$ is a positive integer. We assume that there exists an interaction range $Y \in C_L^a$ such that 
any interaction range $X \in C_L^a$ can be represented in a translation  $X=x+Y$
 with a suitable  $x \in \Lambda_L$. 
The Hamiltonian consists of $m$ terms
\begin{equation}
H(S, g) :=\sum_{a \in A} \sum_{X \in C_L^a} (J_1^a g_X^a +J_0^a) S_X^{i(a)},
\label{Hamil}
\end{equation}
 where a mapping $i : A \rightarrow \{1,2,3\}$ defines a model and $(g_X^a)_{X\in C_L^a, a \in A}$ 
are i.i.d. standard Gaussian random variables or several $g^a$ can be identified to each other.\\

\noindent 
{\it Examples} \\
1. Random  field Heisenberg model \\
$\Lambda_L= {\mathbb Z}^d \cap [1,L]^d$ is a cubic lattice, $A=\{1,2 \}$, $C_L^1=\Lambda_L$,
$C_L^2=\{\{ x,y\}|x,y \in \Lambda_L, |x-y|=1 \}$ is a collection of bonds and $J_0^1=0, J^2_1=0$.
The Hamiltonian is given by
\begin{equation} 
H(S,g)=-J_1^1 \sum_{x \in \Lambda_L} g_x  S_x^z -\sum_{X \in C_L^2} \sum_{i=1}^3 J^2_0 S_X^i.
\end{equation}
\\

2. Random bond Heisenberg model 
\\
For a cubic lattice $\Lambda_L= {\mathbb Z}^d \cap [1,L]^d$ and a collection of bonds  $C_1=\{\{ x,y\}| x,y \in \Lambda_L, |x-y|=1 \}$ 
%and $H_{\rm non} (S)=\sum_{X \in B_L} \sum_{\mu=x,y,z}J g_0 S_X^\mu$ with a real constant $g_0$, the Hamiltonian becomes
\begin{equation}
H(S,g)=- \sum_{X \in C_L^1} \sum_{i=1}^3 (J_1^i g_X ^{i} + J_0^i )S_X^i.
\end{equation}
If $g_X^1 = g_X^2=g_X^3$and $J_r^1=J_r^2=J_r^3$ are identified for all $X \in C_L^1$ and for $r=0,1$, the Hamiltonian is invariant under SU(2) transformation.
\\

3. Other models\\
The Hamiltonian (\ref{Hamil}) contains some other physically interesting models, such as Heisenberg model with random next nearest neighbor
interactions,  and 
with random plaquette  interactions. \\ 

Here, we define Gibbs state for the Hamiltonian.
For a positive $\beta $ and a real number $J$,  the  partition function is defined by
\begin{equation}
Z_L(\beta, J,g) := {\rm Tr} e^{ - \beta H(S,g)},
\end{equation}
where the trace is taken over the Hilbert space ${\cal H}$.
The probability $P_k$ that an energy eigenstate $\phi_k$ with its eigenvalue $E_k$ appears is given by
$$
P_k :=\frac{e^{-\beta E_k}}{Z_L(\beta, J,g) }.
$$     
Let   $f$ be an arbitrary function 
of spin operators.  If $\psi_i$ is  an eigenstate with its eigenvalue $\lambda_i$ of $f$,
the probability that observed value of $f$ is the eigenvalue  $\lambda_i$ is given by
$$
P(f=\lambda_i) = \sum_{k} |(\psi_i, \phi_k)|^2 P_k.
$$ 
Therefore, the  expectation of $f$ in the Gibbs state is given by
\begin{equation}
\langle f(S) \rangle =\sum_{i} \lambda_iP(\lambda_i) =\frac{1}{Z_L(\beta,J,g)}{\rm Tr} f(S)  e^{ - \beta H(S,g)}.
\end{equation}
We define the following functions of  $(\beta,  J) \in [0,\infty) \times {\mathbb R}^{2|A|}$ and randomness
$g=(g_X^a)_{X \in C_L^a, a \in  A}$
\begin{equation}
\psi_L\beta, J,g) := \frac{1}{|\Lambda_L|} \log Z_L(\beta,J,g), \\ 
\end{equation}
$-\frac{L}{\beta}\psi_L(\beta,J,g)$ is called free energy in statistical physics.
%and
We define a function $p_L:[0,\infty) \times {\mathbb R}^{2|A|} \rightarrow {\mathbb R}$ by
\begin{eqnarray}
p_L(\beta,J):={\mathbb E} \psi_L(\beta,J,g) ,
\end{eqnarray}
where ${\mathbb E}$ stands for the expectation of the random variables $(g_X^a)_{X \in C_L^a,a\in A}$.
%The function $p_L(\beta,J)$ is called pressure.
Here, we introduce a fictitious time  $t \in [0,1]$ and define a time evolution of operators with the Hamiltonian.
Let $O$ be an arbitrary self-adjoint operator, and we define an operator valued function  $O(t)$ of $t\in[0,1]$  by
\begin{equation}
  O(t):= e^{-tH} O  e^{tH}.
\end{equation}
Furthermore, we define the  Duhamel expectation of time depending operators 
$ O_1(t_1),  \cdots,  O_k(t_k)$  by
$$
( O_1,  O_2,\cdots,   O_k)_{\rm D} :=\int_{[0, 1]^k} dt_1\cdots dt_k \langle {\rm T}[ O_1(t_1)  O_2(t_2) \cdots  O_k(t_k) ]\rangle,
$$
where the symbol ${\rm T}$ is a multilinear mapping of the chronological ordering.
If we define a partition function with arbitrary self adjoint operators  $O_0, O_1, \cdots, O_k$ and real
numbers $x_1, \cdots, x_k$
$$
Z(x_1,\cdots, x_k) := {\rm Tr} \exp \beta \left[O_0+\sum_{i=1} ^k x_i O_i \right],
$$
the Duhamel expectation of $k$ operators represents
 the $k$-th order derivative of the partition function %$Z(x_1,\cdots, x_k)$ 
 \cite{Cr,GUW,S}
$$\beta^k( O_1,\cdots,  O_k)_{\rm D}=\frac{1}{Z}
\frac{\partial ^k Z}{\partial x_1 \cdots \partial  x_k}.
$$

To study replica symmetry, we define $n$ replicated spin operators $(S_x^{i,\alpha})_{\alpha=1,\cdots, n}$ at each site $x \in \Lambda_L$ and a 
replica symmetric Hamiltonian
\begin{equation}
H(S^1, \cdots, S^n,g):=\sum_{\alpha=1}^n H(S^\alpha, g),
\end{equation}
which  is invariant  under an arbitrary permutation $\sigma \in S_n$  
$$ H(S^1, \cdots, S^n g)=H(S^{\sigma1}, \cdots, S^{\sigma n} , g),$$
as well as the Ising systems.
 The covariance of these operators with the expectation in $g$  for $a \in A$
\begin{equation} {\mathbb E}H_L^a (S^{i(a),\alpha}) H_L^a (S^{i(a), \beta})
- {\mathbb E}(H_L^a (S^{i(a),\alpha})  {\mathbb E}H_L^a (S^{i(a), \beta})
=  |C_L^a| R^a_{\alpha,\beta}
\end{equation}
where  the overlap $R^a_{\alpha,\beta}$  is defined by
$$
R^a_{\alpha,\beta}:=\frac{1}{|C_L^a|} \sum_{X \in C_L^a}S^{i(a),\alpha}_{X}S^{i(a), \beta}_{X}.
$$
For example, in the random field Heisenberg model, this becomes the site overlap operator
$$R^i_{\alpha,\beta}=\frac{1}{|\Lambda_L|} \sum_{x \in \Lambda_L}S^{i, \alpha}_x S^{i,\beta}_x,$$
where we have identified the index $a$ to $i(a)$ for simpler notation.
In the random bond Heisenberg model, it becomes the bond overlap operator
$$
R^i_{\alpha,\beta}=\frac{1}{|B|} \sum_{X \in B}S^{i,\alpha}_{X}S^{i,\beta}_{X}= 
\frac{1}{|B|} \sum_{\{x,y\} \in B} S^{i,\alpha}_{x}  S^{i,\alpha}_ {y}  S^{i, \beta}_{x}   S^{i,\beta}_{y}.
$$
In short range spin glass models, such as  the Edwards-Anderson model \cite{EA} the bond overlap is independent of the site overlap
unlike the Sherrington-Kirkpatrick (SK) model \cite{SK}, where the bond overlap is identical to the square of the site overlap. 
In calculation of thermodynamic quantities
of quantum systems, the following self overlap operator appears quite frequently
$$
R_{1,1}^a:=\frac{1}{|C_L^a|} \sum_{X \in C_L^a} S_X^{i(a),1}(t_1) S_X ^{i(a),1}(t_2)
$$
for $(t_1,t_2) \in [0,1]^2.$ We denote
$$
(R_{1,1}^a)_{\rm D} = \frac{1}{|C_L^a|} \sum_{X \in C_L^a} ( S_X^{i(a),1},  S_X ^{i(a),1})_{\rm D}.
 $$
Note that  $R_{1,1}^a \rightarrow 1$ In the classical limit.
 As we will see later, the expectation values of overlap operators satisfy the quantum mechanically extended Ghirlanda-Guerra identities and Aizenman-Contucci identities (\ref{GG}) which depend on self-overlap operators.

\section{
Self-averaging observables
}
Here, we discuss the self-averaging observables in quantum systems.
Let $O_1, \cdots, O_N$ be a sequence of self-adjoint operators on ${\cal H}$, and we define their mean by
$$
m_N :=\frac{1}{N} \sum_{n=1} ^N O_n. 
$$
Let $\mu$ be an eigenvalue of $m_N$ for arbitrary fixed $(g_X^a)_{X\in C_L^a, a\in A}$.
The probability that the deviation between the observed value $\mu$ of $m_N$ between the expectation  $\langle m_N \rangle $in the Gibbs state is larger than  an arbitrary positive number $c$
 is bounded by the variance  of $m_N$
$$
P(|\mu -\langle m_N \rangle| > c ) \leq \frac{\langle \delta m_N^2\rangle}{c^2}
$$
where  $\delta m_N := m_N-\langle m_N \rangle.$ If the variance $\langle \delta m_N^2\rangle$ 
vanishes in the infinite volume limit, then the observed value differs from the  expectation $\langle m_N \rangle$ rarely.
This implies that  $m_N$ obeys the law of large numbers.
Next, we consider events in a synthesized quantum spin system with quenched disorder.
% given by a random sampling of $(g_X)_{X \in C_L^a, a\in A}$.
A disordered sample synthesizing corresponds to a fixing of random variables $(g_X^a)_{X\in C_L^a, a\in A}$ generated by 
a random sampling. 
We can evaluate the probability that deviation between the observed value $\mu$ and the sample and the Gibbs expectation 
${\mathbb E}\langle m_N \rangle$ is larger than $c>0$ as follows
$$
P(|\mu -{\mathbb E}\langle m_N \rangle| > c ) \leq \frac{{\mathbb E}\langle \Delta m_N^2\rangle}{c^2},
$$
where  $\Delta m_N := m_N-{\mathbb E }\langle m_N \rangle.$  If the variance ${\mathbb E}\langle \Delta m_N^2\rangle$ vanishes in the infinite volume limit, then $m_N$ obeys the law of large numbers. In this case, we can say  that $m_N$ is self-averaging.

Through out this section,  we assume the following general form of the Hamiltonian 
$$
H(O,g):=\sum_{a \in A} \sum_{X \in C_L^a}(J_1^a g_X ^a+J_0^a) O_X^a,
$$
 where each self-adjoint operators $(O_X^a)_{X\in C_L^a,a\in A}$  are confined in $X \in C_L^a$ with $|X| = n_a$ for each $X \in C_L^a$.
 If $X \cap Y=\phi$, the commutator between two operators satisfies
 $$
 [O^{a}_{X}, O^{b}_{Y}] =0.
 $$
 Each $O^a_X$ $({a \in A, X \in C_L^a})$ and the  commutator among any of  them are  bounded
 by  constants independent of the system size $N$.
 We denote 
 $$K^a:=\sup_{\phi \in {\cal H}}\frac{| (\phi,  O_X^a \phi )|}{(\phi,\phi)}.$$
 %We denote its time evolution by$$ H_N ^a(t) := H_L^a( O^a(t)).$$
Here we define a density of a term in the Hamiltonian with the randomness by
\begin{equation}
h_L^a := \frac{1}{|C_L^a|}\sum_{X \in C_L^a}g_X^a O^a_X, 
\label{rand}
\end{equation}
and a deterministic term in the Hamiltonian by 
\begin{equation}
m_L^a := \frac{1}{|C_L^a|}\sum_{X \in C_L^a} O^a_X.
\label{dens}
\end{equation}

Here,  we define two types of deviations of an arbitrary operator $O$ by
$$\delta O:= O -\langle O \rangle, \ \ \  \Delta O := O- {\mathbb E} \langle O \rangle.$$
We prove the following two lemmas for these deviations of 
$$m^a_L=\frac{1}{|C_L^a|}\sum_{X\in C_L^a} O_X^a.$$
%where $O^a_X = S_X^{i(a)}$ for $a \in A$ and $O^0_X = S_X^{i(0),1}S_X^{i(0),2}$.

We assume
the existence of the following  infinite volume limit independent of boundary conditions 
$$
p(\beta,J) = \lim_{L \rightarrow \infty} p_L(\beta, J), \ \ \psi_L(\beta,J) = \lim_{L \rightarrow \infty} \psi_L(\beta, J), 
$$
as proved in \cite{AGL,CGP,CL,I}.

Hereafter, we use a lighter notation $\psi_L(g)$ for $\psi_L(\beta,J,g)$.  
Here we define square root interpolating  random variables $G(\vec u)=(G^a_X(\vec u))_{X \in C_L^a,a \in A}$  for an arbitrary 
vector  $\vec u =(u^a)_{a\in A}  \in [0,1]^{|A|} $by
\begin{equation}
G_X^a(\vec u):= \sqrt{u^a} g^a_X +\sqrt{1-u^a} {g_X^a }',
\end{equation}
where $({g_X^a}')_{a \in C_L^a, a \in A}$ are i.i.d. standard Gaussian random variables.
Then, we define a generating  function $\gamma_L(\vec u)$ of a parameter $\vec u \in [0,1]^{|A|} $ by
\begin{equation}
\gamma_L(\vec u) = {\mathbb E} [{\mathbb E}' \psi_L( G(\vec u) )]^2,
\end{equation}
where  %$|C_L^a|$ dimensional vectors $g=(g_X^a)_{X \in C_L^a}$ and $g'=(g'_i)_{i \in \Lambda_L_L}$
%consist of i.i.d. standard Gaussian variables and 
${\mathbb E }$ and  ${\mathbb E }'$ denote expectation
in $g$ and $g'$, respectively. This generating function $\gamma_L$ is  a generalization of a function introduced by Chatterjee \cite{C}.\\

First,  we present useful lemmas proved in Ref.\cite{I} as Lemma \ref{1}-\ref{cDelta} in the following. 

{\lemma \label{1}  
For any $(\beta, J) \in [0,\infty) \times {\mathbb R}^{2|A|}$, any positive integer
$L$, any positive integer $k$ and any $\vec u_0 \in [0,1]^{|A|}$ whose $a$-th component is $u^a=u_0 < 1$, 
%$\beta, J$, $J_0$ and $L$, %$$Var(\psi_L)=\gamma_L(1)-\gamma_L(0),$$
an upper bound on the $k$-th order partial derivative of the function $\gamma_L$ is given by 
\begin {equation}
\frac{\partial^k \gamma_L}{{\partial u^a}^k} (\vec u_0) \leq \frac{(k-1)! }{(1-u_0)^{k-1}}\frac{(\beta J_1^a K^a)^2 n_a}{|\Lambda_L|}.
\end{equation}
The $k$-th order derivative of $\gamma_L$ is represented in the following
\begin{eqnarray}
\frac{\partial^k \gamma_L}{{\partial u^a}^k} (\vec u) &=& \sum_{X_1 \in C_L^a}  \cdots \sum_{X_k \in C_L^a}
{\mathbb E} \left({\mathbb E}' \frac{\partial^k \psi_L }{\partial g_{X_k}^a \cdots \partial g_{X_1}^a }(G(\vec u))\right)^2 .
%\\&=&\frac{\beta^{2(n-1)} h^{2n}}{|\Lambda_L_L|}  \sum_{i_1 \in \Lambda_L_L}  \cdots \sum_{i_n \in \Lambda_L_L}
%{\mathbb E} ('\langle S_{i_1}; \cdots ; S_{i_n} \rangle_t)^2,
\label{kth}
\end{eqnarray}
for an arbitrary $\vec u \in [0,1]^{|A|}.$
%\noindent Proof. 
} \\

%This can be proved by the same way as in Ref.  \cite{I} . $\Box$\\
% gives an upper bound  on the variance of the function $\psi_L$.
If we define a function $\chi_L:[0,1] \rightarrow {\mathbb R}$ by $\chi_L(s) :=\gamma_L(s,s,\cdots, s)$ for $s \in [0,1]$, then 
Lemma \ref{1} gives the following

{\lemma \label{free}  The variance of 
$ \psi_L$ is bounded from the above as follows
$$
{\mathbb  E} (\psi_L-p_L )^2 =\chi_L(1)-\chi_L(0)  \leq   \frac{(\beta J_1^a K^a)^2 n_a |A|}{|\Lambda_L|}.
$$
\noindent
%Proof.
}
%\begin{eqnarray}
%Var(\psi_L)&=&\chi_L(1)-\chi_L(0)=\int_0^1ds \chi_L'(s)  \leq \chi_L'(1) = \sum_{a\in A} \frac{\partial \gamma_L}{\partial u^a}(1,1,\cdots, 1) %\nonumber \\ &\leq& \sum_{a \in A} \frac{(\beta J_1^a K^a)^2 n_a}{|\Lambda_L|}= \frac{(\beta J_1^a K^a)^2 n_a |A|}{|\Lambda_L|}.\end{eqnarray}
%This completes the proof. $\Box$ \\

{\lemma  \label{delta} 
For  any $a \in  A $ and for $\beta J_1 ^a \neq 0$,  the following quantity is bounded from the above, 
\begin{equation}
 {\mathbb E}  ( {\delta m^a_L }, {\delta m^a_L })_{\rm D} \leq  \frac{K^a}{\beta J_1^a}\sqrt{\frac{1}{n_a|\Lambda_L|}}. 
 \label{a}
\end{equation}
%\noindent Proof. 
}
%This can be proved by the same way as in Ref. \cite{I}. $\Box$\\
%Next we prove the same property of random Hamiltonian density $h^a_L$ as that of $m^a_L$
{\lemma  \label{deltah} 
For any $a \in A$ and any $\beta J^a_1 \neq 0$,  we have 
\begin{equation}
 {\mathbb E}( \delta { h^a_L}, \delta { h^a_L} )_{ \rm D} \leq  \frac{K^a}{\beta J^a_1}\Big(\sqrt{\frac{6}{n_a|\Lambda_L|}} + \frac{1}{n_a|\Lambda_L|}\Big).\label{a1}
\end{equation}
%Proof. 
} 
%This can be proved the same way as in Ref. \cite{I}. $\Box$
\paragraph{Note}  The functions $p(\beta,J),$ $p_L(\beta,J)$ and $\psi_L(\beta,J,g)$ are convex with respect to $\beta$ and $J_r^a (a \in A, r=0,1)$ each.  
This implies that  these functions are differentiable almost everywhere in the coupling constant space $[0,\infty) \times {\mathbb R}^{2|A|}.$
\paragraph{Notation} Let $D$ be the measure zero subset of the coupling constant space
$[0,\infty) \times {\mathbb R}^{2|A|}$ where $p(\beta, J)$ is not differentiable with respect to 
$\beta$ or some $J_r^a$.
{\lemma \label{Delta} 
In the infinite volume limit, the following  is valid
\begin{equation}
\lim _{L \rightarrow \infty} {\mathbb E} \Big( \frac{\partial \psi_L}{\partial J^a_r}- \frac{\partial  p}{\partial J^a_r} \Big)^2
 =0.
 \label{lim2}
\end{equation}
 on  $D^c \subset [0,\infty) \times  {\mathbb R}^{2|A|}$.
 \\
%\noindent Proof.
} 
% Here we regard $\psi_L $ and $p_L$ as functions of a coupling $J_r^a$ for arbitrarily fixed $a \in A$  and $r=0, 1$
%The proof is done by the same way as in Ref. \cite{I} $\Box$

{\lemma  \label{cDelta} In $D^c \subset [0,\infty) \times {\mathbb R}^{2|A|}$, we have  
\begin{equation}
\frac{\partial p}{\partial J^a_r}=\lim_{L \rightarrow \infty}{\mathbb E} \frac{\partial \psi_L}{\partial J^a_r}.
\end{equation}
%Proof.
}
%The above identity (\ref{lim2}) and convexity of the square give 
%\begin{eqnarray}\Big(\lim_{L \rightarrow \infty}{\mathbb E} 
%\frac{\partial \psi_L}{\partial J^a_r} -\frac{\partial p}{\partial J^a_r}\Big)^2  \leq \lim_{L \rightarrow \infty}{\mathbb E} 
%\Big(\frac{\partial \psi_L}{\partial J^a_r} -\frac{\partial p}{\partial J^a_r}\Big)^2 =0, \nonumber  \end{eqnarray}
%since $p$ is differentiable on $D^c$. $\Box$ \\

Next, we prove upper bound on variances of several quantities.    
{\lemma  \label{delta.2} 
For  any $a \in  A $ and for $\beta J_1 ^a \neq 0$, there exists a positive constant $K'$ independent of $L$, such that
\begin{equation}
{\mathbb E}  \langle {\delta m^a_L }^2 \rangle \leq \frac{K^a}{\beta J_1^a}\sqrt{\frac{1}{n_a|\Lambda_L|}}  +\frac{\beta K'}{12 |\Lambda_L|}.
\end{equation}
\noindent
Proof.}
We use Harris'  Bogolyubov type inequality 
between the Duhamel product and the Gibbs expectation of the square of arbitrary self-adjoint operator $O$ \cite{H}
\begin{equation}
( O,  O)_{\rm D} \leq \langle {O}^2\rangle \leq ( O,  O)_{\rm D}  +\frac{\beta}{12} \langle[ O, [H,O]] \rangle,
\label{harris}
\end{equation}
for the above inequality. %If we denote $O_{X,Y,Z} ^{a,b} = [O_X^a,[ O_Z^b,O_Y^a]]$ consisting of bounded spin operators and
We use an indicator $I$ defined by $I[true]=1$ and $I[false]=0$.  Using the Cauchy-Schwalz inequality  and boundedness of operators
$O^a_X$ and their commutatotrs, we have the following
\begin{eqnarray}
{\mathbb E}\langle {\delta m^a_L}^2\rangle &\leq& {\mathbb E}(\delta m^a_L, \delta m^a_L)_{\rm D}  +\frac{\beta}{12|C_L^a|^2} \sum_{X,Y \in C_L^a}\sum_{b\in A} \sum_{Z \in C_L^b} {\mathbb E} g^b_Z \langle [O_X^a,[ O_Z^b,O_Y^a ]] \rangle \nonumber \\
&&\leq {\mathbb E}(\delta m^a_L,\delta m^a_L)_{\rm D}  %\nonumber \\&&
+\frac{\beta}{12|C_L^a|^2} \sum_{b\in A} \sum_{Z \in C_L^b} \sum_{X,Y \in C_L^a}
\sqrt{ {\mathbb E} (g^b_Z)^2 {\mathbb E}\langle [O_X^a,[ O_Z^b,O_Y^a ]] \rangle^2 } \nonumber \\
&&\leq{\mathbb E}(\delta m^a_L,\delta m^a_L)_{\rm D}  %\nonumber \\
+\frac{\beta}{12|C_L^a|^2} \sum_{b\in A} \sum_{Z \in C_L^b} \sum_{X,Y \in C_L^a} I[X \cap (Y\cup Z)\neq \phi]I[Y\cap Z\neq \phi]
 K \nonumber  \\
%&\leq&{\mathbb E}(m^a_L m^a_L)_{\rm D}  +\frac{\beta}{12|C_L^a|^2} \sum_{b\in A}  \sum_{Z \in C_L^b} \sum_{X,Y \in C_L^a} I[X \cap Y\cap Z\neq \phi]{\mathbb E}\sqrt{\langle { O_{X,Y,Z} ^{a,b} }^2\rangle \langle {O_Z^b}^2  \rangle } \nonumber  \\
&&\leq {\mathbb E}(\delta m^a_L,\delta m^a_L)_{\rm D}  +\frac{\beta K'}{12 |\Lambda_L|},
\label{b}
\end{eqnarray}
where $K$ and $K'$ are positive constant  independent of $L$. 
Therefore, this and (\ref{a})
%\begin{equation}{\mathbb E}\langle {\delta m^a_L}^2 \rangle \leq \frac{K^a}{\beta J_1^a}\sqrt{\frac{1}{|\Lambda_L|}}  +\frac{\beta K'}{12 |\Lambda_L|}.\label{c}\end{equation}
complete the proof. 
$\Box$
\\
%Since we have $C |\Lambda_L_L| \leq |C_L | \leq C'| \Lambda_L_L|$ for some $C,C'$ independent of $L$,  thisgives the limit (\ref{a}).
{\lemma  \label{delta2} 
For any $a \in A$ and any $\beta J^a_1 \neq 0$,  there exists a positive number $K'$
we have 
\begin{equation}
{\mathbb E}\langle \delta { h^a_L}^2 \rangle \leq \frac{K^a}{\beta J^a_1}\Big(\sqrt{\frac{6}{n_a|\Lambda_L|}} + \frac{1}{n_a|\Lambda_L|}\Big)  +\frac{\beta K'}{12 |\Lambda_L|}.
\label{a2}
\end{equation}
Proof. }
As well as in the proof of Lemma \ref{delta.2}, we use the inequality (\ref{harris})  and the Cauchy-Schwalz inequality 
\begin{eqnarray}
%&&
{\mathbb E}\langle {h^a_L}^2\rangle  %\nonumber \\
&\leq& {\mathbb E}(h^a_L, h^a_L)_{\rm D}  +\frac{\beta}{12|C_L^a|^2} \sum_{X,Y \in C_L^a}\sum_{b\in A}  \sum_{Z \in C_L^b} {\mathbb E}g^a_X g^a_Y  g^b_Z \langle [O_X^a,[ O_Z^b,O_Y^a ]] \rangle \nonumber \\
&\leq& {\mathbb E}(h^a_L, h^a_L)_{\rm D}  +%\nonumber \\&+&
\frac{\beta}{12|C_L^a|^2} \sum_{b\in A} \sum_{Z \in C_L^b} \sum_{X,Y \in C_L^a}% I[X \cap (Y\cup Z)\neq \phi]I[Y\cap Z\neq \phi] 
\sqrt{{\mathbb E}(g^a_X g^a_Y  g^b_Z)^2  {\mathbb E} \langle [O_X^a,[ O_Z^b,O_Y^a ]] \rangle ^2} \nonumber \\
&\leq&{\mathbb E}(h^a_L, h^a_L)_{\rm D}  %\nonumber \\
+\frac{\beta}{12|C_L^a|^2} \sum_{b\in A}  \sum_{Z \in C_L^b} \sum_{X,Y \in C_L^a}I[X \cap (Y\cup Z)\neq \phi]I[Y\cap Z\neq \phi]  K \nonumber  \\
&\leq&{\mathbb E}(h^a_L, h^a_L)_{\rm D}  +\frac{\beta K'}{12 |\Lambda_L|},  \label{b} 
\end{eqnarray}
Therefore, this and (\ref{a1}) 
%\begin{equation}{\mathbb E}\langle {h^a_L}^2 \rangle \leq \frac{K^a}{\beta J^a_1}\Big(\sqrt{\frac{6n_a}{|\Lambda_L|}} 
%+ \frac{1}{|\Lambda_L|}\Big)  +\frac{\beta K'}{12 |\Lambda_L|}.\label{c2}\end{equation} This inequality
complete the proof. $\Box$\\

Aizenman, Greenbatt and Lebowitz proved the rounding effect of disordered interactions which means that $p$ is differentiable with respect to $J_0^a$ for $J_1^a \neq 0$ in $d \leq 2$ for a system with discrete symmetry
 and  in $d \leq 4$  for a system with a continuous symmetry
\cite{AGL,GAL}. Their proof is an extension of the differentiability of the free energy density for classical systems proved by
Aizenman and Wehr \cite{AW} to the quantum systems.
In dimensions $d$ larger than the critical dimensions, there is a possibility that the function $p$ is not differentiable on a measure zero subset of the coupling constant space, where the order parameter can be discontinuous  and  the first order phase transition occurs.  In this case, the Gibbs state is not unique. The discontinuity appears at $J_0^a=0$ most likely.\\

%For example, in the random field Ising model, spontaneous magnetization exists in three dimensions at sufficiently low temperatures \cite{GK}. 
Note the relations
$$\beta  \langle m_L^a \rangle = \frac{\partial \psi_L}{\partial J_0^a}, \hspace{.5cm}\beta  \langle  h_L^a \rangle =\frac{\partial \psi_L}{\partial J_1^a}, \hspace{.5cm} \beta{\mathbb E}  \langle m_L^a \rangle = \frac{\partial p_L}{\partial J_0^a}, \hspace{.5cm}
\beta  {\mathbb E} \langle  h_L^a\rangle =\frac{\partial p_L}{\partial J_1^a}. $$
 
If $\frac{\partial p}{\partial J^a_r}$ has a discontinuity at ${J^a_r}'$,  differentiability of $p$ almost every where around  ${J_r^a}'$
gives
$$
\lim_{x\downarrow {J^a_0}'}  \frac{\partial p}{\partial J^a_0}(x) = \lim_{x\downarrow {J^a_0}'} \lim_{L\uparrow \infty} \beta {\mathbb E} \langle  m_L^a \rangle(x), \ \ \  \ \ \ \ \  \lim_{x\uparrow {J^a_0}'}  \frac{\partial p}{\partial J^a_0}(x) = \lim_{x\uparrow {J^a_0}'} \lim_{L\uparrow \infty} \beta {\mathbb E} \langle  m_L^a \rangle(x).
$$
Also integration by parts enables us to calculate the following
$$
\lim_{x\downarrow J{^a_1}'}  \frac{\partial p}{\partial J^a_1}(x) = \lim_{x\downarrow{ J^a_1}'} \lim_{L\uparrow \infty} \beta \sum_{X \in C_L^a} {\mathbb E}g_X^a \langle  O_X^a \rangle(x)= \lim_{x\downarrow {J^a_1}'} \lim_{L\uparrow \infty} \beta^2J_1^a \sum_{X \in C_L^a} {\mathbb E}[
(  O_X^a,  O_X^a )_{\rm D}(x)- \langle  O_X^a \rangle(x) ^2], $$
$$ \lim_{x\uparrow {J^a_1}'}  \frac{\partial p}{\partial J^a_1}(x) = \lim_{x\uparrow{ J^a_1}'} \lim_{L\uparrow \infty} \beta \sum_{X \in C_L^a} {\mathbb E}g_X^a \langle  O_X^a \rangle(x)= \lim_{x\uparrow {J^a_1}'} \lim_{L\uparrow \infty} \beta^2J_1^a \sum_{X \in C_L^a} {\mathbb E}[
(  O_X^a,  O_X^a )_{\rm D}(x)- \langle  O_X^a \rangle(x) ^2].
$$
These identities imply that the order parameter becomes discontinuous at the non-differentiable point of $p$.\\

Now, we prove 
that $m_\infty^a$ and $h_\infty^a$  are self-averaging on $D^c$. \\

{\theorem  \label{MT1} 
 For $\beta J^a_1 \neq 0$,
variances of the density  defined by $m_L^a:= \frac{1}{|C_L^a|} \sum_{X \in C_L } O_X ^{a}$ and that of density of the Hamiltonian with randomness $h_L^a:=\frac{1}{|C_L^a|} \sum_{X \in C_L }g_X^a O_X ^{a}$ 
for an arbitrary $a \in A$ %defined by (\ref{dens}) 
vanish  in the infinite volume limit
\begin{eqnarray}
%\lim_{L \rightarrow \infty } {\mathbb E}\langle(\psi_L(\beta, \vec J, \vec g ) -p(\beta, \vec J) )^2 \rangle =0, %\nonumber \\
&&\lim_{L \rightarrow \infty } {\mathbb E}
\langle (m^a_L -  {\mathbb E} \langle m^a_L \rangle)^2 \rangle =0, \label{limit2}  \\%\nonumber 
&&\lim_{L \rightarrow \infty } {\mathbb E}
\langle (h^a_L -  {\mathbb E} \langle h^a_L \rangle)^2 \rangle =0,  \label{limit1} 
\end{eqnarray}
 on
$D^c \subset [0,\infty) \times {\mathbb R}^{2|A| }$.  \\

Proof }
We calculate the following variance of $m^a_L$, 
\begin{eqnarray}
&&\lim_{L \rightarrow \infty}[ {\mathbb E} \langle {m_L^a}^2 \rangle -( {\mathbb E} \langle {m_L^a} \rangle)^2]=
\lim_{L \rightarrow \infty}[ {\mathbb E} \langle {m_L^a}^2 \rangle 
- {\mathbb E} \langle {m_L^a} \rangle^2]+\lim_{L \rightarrow \infty} [{\mathbb E} \langle {m_L^a} \rangle^2
-( {\mathbb E} \langle {m_L^a} \rangle \rangle)^2] \nonumber \\
&&=
\lim_{L \rightarrow \infty} {\mathbb E} \langle( {m_L^a} 
- \langle {m_L^a} \rangle )^2\rangle+\lim_{L \rightarrow \infty} {\mathbb E} (\langle {m_L^a} \rangle
- {\mathbb E} \langle {m_L^a} \rangle )^2 =0.
\label{correq}
\end{eqnarray}
 We have used Lemma \ref{delta.2} for the first term and Lemma \ref{Delta} for the second term in the last line.
We calculate the following variance of $h^a_L$, as well.
\begin{eqnarray}
&&\lim_{L \rightarrow \infty}[ {\mathbb E} \langle {h_L^a}^2 \rangle -( {\mathbb E} \langle {h_L^a} \rangle)^2]=
\lim_{L \rightarrow \infty}[ {\mathbb E} \langle {h_L^a}^2 \rangle 
- {\mathbb E} \langle {h_L^a} \rangle^2]+\lim_{L \rightarrow \infty} [{\mathbb E} \langle {h_L^a} \rangle^2
-( {\mathbb E} \langle {h_L^a} \rangle \rangle)^2] \nonumber \\
&&=
\lim_{L \rightarrow \infty} {\mathbb E} \langle( {h_L^a} 
- \langle {h_L^a} \rangle )^2\rangle+\lim_{L \rightarrow \infty} {\mathbb E} (\langle {h_L^a} \rangle
- {\mathbb E} \langle {h_L^a} \rangle )^2 =0.
\label{correq2}
\end{eqnarray}
We have used Lemma \ref{delta2} for the first term and Lemma \ref{Delta} for the second term in the last line.
Therefore, $m_L^a$ and $h^a_L$ are self-averaging. 
$\Box$

{\theorem  \label{MT3} The following conditions {\rm (I)} and {\rm (II)}  are equivalent for almost all sequences of random variables $(g_X^a)$.\\

\noindent
{\rm (I)}  There exists $b\in A$, such that
\begin{equation}
 % {\cal S}_{K} := \{(g^a_X)\in \prod_{a \in A} {\mathbb R}^{|C_K^a|}; 
 \lim_{L \rightarrow \infty } %{\mathbb E}
\langle (m_L^b -   \langle m_L^b \rangle)^2 \rangle \neq  0,
\label{I}
\end{equation}
  at $(J_0^b,J_1^b)=(0,0)$.\\
 
\noindent
{\rm (II)} 
There exists $b \in A$, such that
\begin{equation}{\displaystyle
 \lim_{J_0^b\rightarrow 0}   \lim_{J_1^b\rightarrow 0} \lim_{L \rightarrow \infty} %{\mathbb E}
 \langle({m_L^a} -\langle{m_L^b} \rangle)^2\rangle \neq
 \lim_{L \rightarrow \infty} 
 \lim_{J_0^b \rightarrow 0} \lim_{J_1^b\rightarrow 0}  %{\mathbb E}
 \langle ({m_L^b} -\langle{m_L^b} \rangle)^2\rangle.}
\label{II}
\end{equation}
%for  almost all  $(g^a_X)\in {\cal S}_{K}$. \\

\noindent
Proof.}  
 Lebesgue's dominated convergence theorem guarantees a commutativity  between any limit operations  and the sample expectation ${\mathbb E}$, since $\langle (m_L^b)^k \rangle$ is bounded by a constant independent of $L$ for any positive $k$. Therefore,
we evaluate  the variance of $m_L^b$
\begin{eqnarray}
%&&
{ \mathbb E}  %I[(g^b_X) \in {\cal S}_{K} ] 
\lim_{J_0^b \rightarrow 0} \lim_{J_1^b \rightarrow 0}\lim_{L \rightarrow \infty }
\langle (m_L^b-\langle m_L^b\rangle  )^2 \rangle%\nonumber \\
%&& =\lim_{J_0^b \rightarrow 0} \lim_{J_1^b \rightarrow 0} \lim_{L \rightarrow \infty }{ \mathbb E} I[(g^b_X) \in {\cal S}_K]  \langle ( m_L^b-\langle m_L^b\rangle  )^2  \rangle \nonumber \\&& \leq
=\lim_{J_0^b \rightarrow 0} \lim_{J_1^b \rightarrow 0} \lim_{L \rightarrow \infty} { \mathbb E}\langle  (m_L^b - \langle m_L^b\rangle )^2\rangle=0.
\nonumber
\end{eqnarray}
The final term vanishes in the coupling constant space $[0,\infty) \times {\mathbb R}^{2|A|}$ by Lemma \ref{delta.2}.  
The positive semi-definiteness of $(m_L^b -
\langle m_L^b\rangle  )^2$  gives that the left hand side in (\ref{II}) vanishes
$$
\lim_{J_0^b \rightarrow 0}  
\lim_{J_1^b \rightarrow 0}  
\lim_{L \rightarrow \infty } \langle
(m_L^b -\langle m_L^b\rangle
)^2 \rangle=0,
$$
for almost all $(g_X^a)$.
%for almost all  $(g_X^a)_{X\in C_K^a} \in \prod_{a \in A} {\mathbb R}^{|C_K^a|}$.  
The continuity of the function of $(J_0^b,J_1^b)$ for an arbitrary finite $L$ guarantees the equivalence between positive variance and non-commutativity of limits for almost all $(g_X^a)$. 
This completes the proof.
$\Box$ 
\\

Here, we apply Theorem \ref{MT3} to the symmetry breaking phenomena in quantum systems with Gaussian disorder.
Assume that the Hamiltonian has a certain symmetry at $(J_0^a,J_1^a)=(0,0)$,  and order parameter $m_L^a$ breaks this symmetry.
%In Theorem \ref{MT3}, (I) implies finite variance 
%in the symmetric Gibbs state and (II) implies spontaneous symmetry breaking.
Theorem \ref{MT3} claims the equivalence between the finite variance (I) in symmetric Gibbs state and the spontaneous symmetry breaking (II).
As in quantum spin systems without disorder, the existence of spontaneous symmetry breaking can be detected by the variance of the corresponding  order parameter calculated in the symmetric Gibbs state. 
The Koma-Tasaki theorem is important for the Heisenberg quantum spin model with SU(2) symmetry.
 This theorem claims that the finite long range order  
$$
\lim_{L \rightarrow\infty}\langle %(
{m^i_L%-\langle m^i_L\rangle )
}^2 \rangle \neq 0,
$$
for the order parameter $m^i_L :=\frac{1}{|\Lambda_L|} \sum_{x \in \Lambda_L } S_x^i$
in the symmetric Gibbs state guarantees 
a finite spontaneous magnetization with an infinitesimal symmetry  breaking field \cite{KT}. 
This is a quantum mechanical extension of Griffiths' theorem \cite{Gff}.
Since $\langle m^i_L \rangle=0$  
in the symmetric Gibbs state in this case, 
 the finite variance implies the long range order.
Although the finite variance of the order parameter seems to violate the law of large numbers, this is just apparent. In this case, one of 
the symmetry breaking Gibbs states obtained by applying the infinitesimal symmetry breaking field
realizes instead of the unstable symmetric one. 
Theorem \ref{MT3} claims the same conclusion for spontaneous symmetry breaking phenomena
in  disordered quantum spin systems as that of the Koma-Tasaki theorem. 
% The condition (I) in Theorem \ref{MT3} can be regarded as long range order. 
If the variance of $m^b_L$ becomes finite in the symmetric Gibbs state, 
one of the symmetry breaking Gibbs states should realize actually instead of the unstable symmetric one
and  the order parameter $m^b_L$ obeys the law of large numbers. \\

Next, we present equivalence between non-self-averaging order parameter in symmetric Gibbs state and 
non-commutativity of the infinite volume limit and the symmetric limit.
 Therefore, this implies that a violation of the self-averaging in the 
symmetric calculation should be raised by a spontaneous symmetry breaking.

{\theorem \label{MT4} 
The following conditions {\rm (III)} and {\rm (IV)}  are equivalent for almost all sequence of random variables $(g_X^a)$ on 
$D^c\subset [0,\infty) \times {\mathbb R}^{2|A|}$.\\

\noindent
{\rm (III)}  There exists $b\in A$, such that 
\begin{equation}
 % {\cal S}_{K} := \{(g^b_X)\in \prod_{a \in A} {\mathbb R}^{|C_K^b|}; 
 \lim_{L \rightarrow \infty } 
\langle (m_L^b -  {\mathbb E} \langle m_L^b \rangle)^2 \rangle \neq  0,
\label{III}
\end{equation}
 at $(J_0^b,J_1^b)=(0,0)$.\\

\noindent
{\rm (IV)} 
There exists $b \in A$, such that
\begin{equation}{\displaystyle
 \lim_{J_0^b\rightarrow 0}   \lim_{J_1^b\rightarrow 0} \lim_{L \rightarrow \infty}\langle({m_L^b} -{\mathbb E}\langle{m_L^b} \rangle)^2\rangle \neq
 \lim_{L \rightarrow \infty} 
 \lim_{J_0^b \rightarrow 0} \lim_{J_1^b\rightarrow 0} \langle ({m_L^b} -{\mathbb E} \langle{m_L^b} \rangle)^2\rangle.}
\label{IV}
\end{equation}
%for  almost all  $(g^b_X)\in {\cal S}_{K}$. \\

\noindent
Proof. } 
 As  in the proof of Theorem \ref{MT3}, we have the commutativity between any limit operations and the sample expectation ${\mathbb E}$.
If we use Lemma  \ref{MT1}, we obtain 
\begin{eqnarray}
%&&
{ \mathbb E}%I[(g^b_X)_{X \in C_K^b} \in {\cal S}_K ]
\lim_{J_0^b \rightarrow 0}\lim_{J_1^b \rightarrow 0} \lim_{L \rightarrow \infty }\langle (m_L^b-{\mathbb E}\langle m_L^b\rangle )^2 \rangle 
%\nonumber \\ && =\lim_{J_0^b \rightarrow 0}\lim_{J_1^b \rightarrow 0} \lim_{L \rightarrow \infty }{ \mathbb E}I[(g^b_X)_{X \in C_K^b} \in {\cal S}_{K} ]\langle ( m_L^b-{\mathbb E}\langle m_L^b\rangle  )^2  \rangle \nonumber \\&& \leq
= \lim_{J_0^b \rightarrow 0}\lim_{J_1^b \rightarrow 0} \lim_{L \rightarrow \infty }{ \mathbb E}\langle ( m_L^b-{\mathbb E}
\langle m_L^b\rangle  )^2  \rangle  =0, \nonumber
\end{eqnarray}
on  $D^c$.  
The positive semi-definiteness of $(m_L^b -{\mathbb E}
\langle m_L^b\rangle  )^2$  gives
$$
\lim_{J_0^b \rightarrow 0}\lim_{J_1^b \rightarrow 0}\lim_{L \rightarrow \infty } \langle (m_L^b-{\mathbb E}
\langle m_L^b\rangle  )^2 \rangle=0, 
$$
for almost all $(g_X^a)$. 
Therefore, the violation of the self-averaging  
$$
\lim_{L \rightarrow \infty } \langle (m_L^b-{\mathbb E}
\langle m_L^b\rangle  )^2 \rangle\neq 0
$$
at $(J_0^b,J_1^b)=(0,0)$
is equivalent to  the following non-commutativity
$$
\lim_{L \rightarrow \infty } \lim_{J_0^b \rightarrow 0}\lim_{J_1^b \rightarrow 0}\langle (m_L^b-{\mathbb E}
\langle m_L^b\rangle  )^2 \rangle \neq
\lim_{J_0^b \rightarrow 0}\lim_{J_1^b \rightarrow 0}\lim_{L \rightarrow \infty } \langle (m_L^b-{\mathbb E}
\langle m_L^b\rangle  )^2 \rangle,
$$
for almost all $(g_X^a)$ on  $D^c$.
This completes the proof. 
$\Box$

\section{The Ghirlanda-Guerra type identities for quantum systems}
The following limit given in Theorem \ref{MT1}
$$
\lim_{L \rightarrow \infty} {\mathbb E} \langle {\Delta h^a_L} ^2\rangle=0,
$$
 enables us to derive quantum mechanically extended identities of the  Ghirlanda-Guerra type
 which differ  from those obtained in Ref.\cite{I}  by the limit
$$
\lim_{L \rightarrow \infty} {\mathbb E} ( {\Delta h^a_L},  {\Delta h^a_L} )_{\rm D}=0.
$$

{\theorem  \label{MT2}
For $\beta J_1^a \neq 0$, on $D^c \subset [0,\infty) \times {\mathbb R}^{2|A|}$, 
for an arbitrary bounded function $f$ of $n$ replicated spin operators, 
the following identity is valid
\begin{eqnarray}
\lim_{L \rightarrow \infty} \Big[&&\frac{1}{|C_L^a|} \sum_{X\in C_L^a}\sum_{\alpha=1} ^n {\mathbb E}(S_X^{i(a),\alpha},S_X^{i(a),1}f)_{\rm D} -n {\mathbb E} \langle R_{1,n+1}^a f \rangle  \nonumber \\
&& \ + \ {\mathbb E}\langle R_{1,2}^a \rangle {\mathbb E}\langle f \rangle
 - {\mathbb E} (R_{1,1}^a)_{\rm D} {\mathbb E}\langle
f \rangle \Big]=0  
\label{QGG}
\end{eqnarray}
Proof. } 
 Theorem \ref{MT1} gives 
\begin{equation}
\lim_{L \rightarrow \infty} {\mathbb E} \langle {\Delta h^a_L}^2 \rangle =0.
\label{limit}
\end{equation}
The Cauchy-Schwarz inequality and the boundedness of $f$ impliy 
$$
|{\mathbb E}\langle  \Delta h^a_L f\rangle| \leq \sqrt{ {\mathbb E}\langle  {\Delta h^a_L}^2 \rangle {\mathbb E}\langle  f^2 \rangle}  \rightarrow 0,
$$
in the infinite volume limit.
The left hand side can be calculated using integration by parts.
\begin{eqnarray}
&&\frac{1}{|C_L^a|}\sum_{X \in C_L^a}{\mathbb E}  g_X^a \langle   S_X^{i(a)} f\rangle 
=\frac{1}{|C_L^a|} \sum_{X \in C_L^a} {\mathbb E} \frac{\partial }{\partial g_X^a} \langle S_X^{i(a)} f\rangle \nonumber \\
&&= \frac{\beta J_1^a}{|C_L^a|} \sum_{X \in C_L^a} [\sum_{\alpha=1}^n {\mathbb E}(S_X^{i(a),1 }, S_X^{i(a),\alpha} f)_{\rm D} -n  {\mathbb E}\langle  S^{i(a)}_X \rangle  \langle   S^{i(a),1}_X f \rangle ]  \nonumber \\
%&&= \frac{\beta J_1^a}{|C_L^a|} \sum_{X \in C_L^a}[ \mathbb E}(S_X^{i(a),1 }, S_X^{i(a),1}f)_{\rm D} + {\sum_{\alpha=2}^n {\mathbb E}\langle S_X^{i(a),1 } S_X^{i(a),\alpha}f\rangle-n {\mathbb E} \langle   S^{i(a),n+1}_X \rangle \langle S^{i(a),1}_X f \rangle  ]  \nonumber \\
&&=\beta J_1^a \Big[\frac{1}{|C_L^a|} \sum_{X \in C_L^a} \sum_{\alpha=1}^n {\mathbb E}(S_X^{i(a),\alpha}, S_X^{i(a),1} f)_{\rm D}-n {\mathbb E}\langle R^a_{1,n+1} f \rangle \Big] 
\end{eqnarray}
Substituting  $f=1$ to the above,  we have
\begin{eqnarray}
&&\frac{1}{|C_L^a|}\sum_{X \in C_L^a}{\mathbb E}  g_X^a \langle S_X^{i(a)} \rangle 
%=\frac{1}{|C_L^a|} \sum_{X \in C_L^a} {\mathbb E} \frac{\partial }{\partial g_X^a} \langle S_X^{i(a)} \rangle 
\nonumber \\
%&&= \frac{\beta J_1^a}{|C_L^a|} \sum_{X \in C_L^a} \sum_{\alpha=1}^n {\mathbb E}[(S_X^{i(a),1 }, S_X^{i(a),\alpha})_{\rm D} -\langle f S^{i(a),\alpha}_X \rangle  \langle   S^{i(a),\alpha}_X \rangle ]  \nonumber \\
%&&= \frac{\beta J_1^a}{|C_L^a|} \sum_{X \in C_L^a} \mathbb E}[(S_X^{i(a),1 }, S_X^{i(a),1})_{\rm D} + {\sum_{\alpha=2}^n\langle fS_X^{i(a),1 } S_X^{i(a),\alpha}\rangle-\sum_{\alpha=1}^n\langle f S^{i(a),\alpha}_X \rangle  \langle   S^{i(a),n+1}_X \rangle ]  \nonumber \\
&&=\beta J_1^a \Big[\frac{1}{|C_L^a|} \sum_{X \in C_L^a} \mathbb E}(S_X^{i(a),1 }, S_X^{i(a),1})_{\rm D} + {\sum_{\alpha=2}^n { \mathbb E}\langle R_{1,\alpha }\rangle-n { \mathbb E}\langle R^a_{1,n+1} \rangle \Big] \nonumber \\
&&=\beta J_1^a[{ \mathbb E}(R_{1,1}^a )_{\rm D} - { \mathbb E}\langle R^a_{1,2} \rangle ]
\end{eqnarray}
From the above two identities, we have 
\begin{eqnarray}
{\mathbb E} \langle \Delta h^a_L f \rangle
%=\frac{1}{|C_L^a|}\sum_{X \in C_L^a}\Big[ {\mathbb E}  g_X^a \langle   S_X^{i(a)}f  \rangle 
%- {\mathbb E}  g_X^a \langle S_X^{i(a)} \rangle  {\mathbb E}\langle f \rangle \Big] \nonumber \\&&
= 
\beta J_1^a \Big[ &&\frac{1}{|C_L^a|} \sum_{X \in C_L^a} \sum_{\alpha=1}^n {\mathbb E}(S_X^{i(a),\alpha }, S_X^{i(a),1}f)_{\rm D}  
-n {\mathbb E}\langle R^a_{1,n+1} f  \rangle \nonumber \\ 
&&-({ \mathbb E}(R_{1,1}^a )_{\rm D} - { \mathbb E}\langle R^a_{1,2} \rangle) {\mathbb E } \langle f \rangle \Big],
\end{eqnarray}
Therefore, we obtain the given identity (\ref{QGG}) $\Box$ \\

Now, we derive the relation between two kinds of variance 
${\mathbb E}\langle {\Delta R_{1,2}^a }^2\rangle $ and $ {\mathbb E}\langle {\delta R_{1,2}^a }^2 \rangle$, 
where $\Delta R_{1,2}^a := R_{1,2} ^a -{\mathbb E}\langle
R_{1,2}^a \rangle$ and $\delta R_{1,2}^a := R_{1,2} ^a -\langle
R_{1,2}^a \rangle.$  
 
For $n=2, f=R^a_{1,2} $ Theorem \ref{MT2} gives
\begin{eqnarray}
\lim_{L \rightarrow \infty} \Big[ &&\frac{1}{|C_L^a|} \sum_{X \in C_L^a} {\mathbb E}(S_X^{i(a),2},S_X^{i(a),1}R_{1,2}^a)_{\rm D}
 -2{\mathbb E} \langle R_{1,3}^a R_{1,2}^a \rangle + 
({\mathbb E}\langle R_{1,2}^a \rangle )^2 \nonumber \\&&
+
\frac{1}{|C_L^a|} \sum_{X \in C_L^a} {\mathbb E}(S_X^{i(a),1},S_X^{i(a),1}\Delta R_{1,2}^a)_{\rm D}
 \Big]=0.
\label{R12}
\end{eqnarray}

For  $n=3, f=R_{2,3}^a$, Theorem \ref{MT2} gives 
\begin{eqnarray}
\lim_{L \rightarrow \infty}\Big[ 2{\mathbb E}( R_{1,2}^a R_{2,3}^a)_{\rm D}
%{\mathbb E}\langle R_{1,3}^a R_{2,3}^a \rangle
- 3{\mathbb E} \langle R_{1,4}^a R_{2,3}^a \rangle &&+ 
{\mathbb E}\langle R_{1,2}^a \rangle {\mathbb E}\langle R_{2,3}^a \rangle \nonumber \\&&
+
%\frac{1}{|C_L^a|} \sum_{X\in C_L^a} {\mathbb E}(S_X^{i(a),1},S_X^{i(a),1}R_{2,3} ^a)_{\rm D}
 {\mathbb E} (R_{1,1}^a)_{\rm D} \langle \Delta
R_{2,3}^a \rangle \Big]=0.  
\label{R23}
\end{eqnarray} 
We have used the replica symmetry, since we calculate above terms in the replica symmetric expectation
The identities (\ref{R12}) and (\ref{R23}) give
\begin{eqnarray}
&&\lim_{L \rightarrow \infty} \Big[2({\mathbb E}\langle R_{1,2}^a \rangle )^2- 3{\mathbb E} \langle R_{1,2}^a \rangle^2
+\frac{1}{|C_L^a|} \sum_{X \in C_L^a} {\mathbb E}(S_X^{i(a),2},S_X^{i(a),1} R_{1,2}^a)_{\rm D} 
\\&&
+
\frac{1}{|C_L^a|} \sum_{X \in C_L^a} {\mathbb E}(S_X^{i(a),1},S_X^{i(a),1}\Delta R_{1,2}^a)_{\rm D}+ {\mathbb E}(R_{1,1} ^a)_{\rm D} \langle \Delta R_{1,2}\rangle+2{\mathbb E}( R_{1,2}^a R_{1,3}^a)_{\rm D} -2{\mathbb E} \langle R_{1,2}^a R_{1,3}^a \rangle\nonumber 
 \Big]=0.
\end{eqnarray} 
This identity has not been very useful so far.
In the classical limit, four terms in the second line vanish, and we obtain
$$
\lim_{L \rightarrow \infty} \Big[2({\mathbb E}\langle R_{1,2}^a \rangle )^2- 3{\mathbb E} \langle R_{1,2}^a \rangle^2
+{\mathbb E} \langle {R_{1,2}^a}^2 \rangle\Big]=0,
$$
which yields
$$2\lim_{L \rightarrow \infty} {\mathbb E}\langle {\Delta R_{1,2}^a }^2 \rangle=
3\lim_{L \rightarrow \infty} {\mathbb E}\langle {\delta R_{1,2}^a }^2 \rangle.
$$
If the right hand side vanishes, then the left hand side vanishes.
In quantum systems however,
we cannot judge whether or not $\lim_{L \rightarrow \infty} {\mathbb E}\langle {\Delta R_{1,2}^a }^2 \rangle$ vanishes even if 
$\lim_{L \rightarrow \infty} {\mathbb E}\langle {\delta R_{1,2}^a }^2 \rangle=0.
$

Note that another relation (\ref{limit2})
in Theorem \ref{MT1} yields the following identity for an arbitrary bounded function $f$ of spin operators   
$$
{\mathbb E} \langle \Delta m_L^a  f \rangle=0.
$$  
This should be useful some times.

\section{Spontaneous replica symmetry breaking}
 Here, we study replica symmetry breaking phenomena applying Theorem \ref{MT3} and \ref{MT4} to the overlap operator
 %To this end, we consider the following Hamiltonian of $n$ replicated spin operators with inter replica coupling   
%$$H(S^1, \cdots, S^n, g)+H_{\rm int} (S^1,S^2,g^0).$$ To study an overlap 
$$R_{1,2} ^c:= \frac{1}{|C_L^c|} \sum_{X \in C_L^c} S_X^{i(c),1} S_X^{i(c),2},$$
for an arbitrary fixed $c \in A$. We extend the coupling constant space $[0, \infty) \times {\mathbb R}^{2|A|}$ 
to $[0, \infty) \times {\mathbb R}^{2(|A|+1)}$ with  new coupling constants $(J^0_0,J^0_1)$ to introduce 
inter replica coupling.  We define a total Hamiltonian by
\begin{equation}
H_{\rm tot}(S^1, \cdots, S^n, g, g^0) := H(S^1, \cdots, S^n,g)+H_{\rm int}(S^1,S^2,g^0),
%\sum_{a \in A*} \sum_{X \in C_L^a} (J_1^a g_X^a +J_0^a) O_X^{a},
\label{Hamil2}
\end{equation}
where
\begin{eqnarray}
&&H(S^1, \cdots, S^n, g)=\sum_{\alpha =1} ^n \sum_{a \in A}\sum_{X \in C_L^a}  (J^a_1 g^a_X+J^a_0) S_X^{i(a),\alpha},  \nonumber \\
&&H_{\rm int} (S^1,S^2,g^0) = \sum_{X\in C_L^0}(J^0_1 g^0_X+J^0_0) S_X^{i(0),1} S_X^{i(0),2}, \label{inthamil}
\end{eqnarray}
with $C_L^0=C_L^c$ and $i(c)=i(0)$. 
Note that the inter replica coupling breaks the replica symmetry of the Hamiltonian.
%where $A*=\{ 0 \} \cup A$ and $O_X^{0} = S_X^{i,\alpha}S_X^{i,\beta}$ and $O_X^{a}=$
We can apply all obtained theorems to this system,  since the operators
$$O^{a}_X:= S_X^{i(a)}$$ for $a \in A$ and
$$O^0_X := S_X^{i(c),1} S_X^{i(c),2}$$ 
for $c \in A$  and $ X \in C_L^0$ 
are bounded by
$$
\sup_{\phi \in {\cal H} } \frac{ |( \phi ,S_X^{i(a)} \phi)|}{(\phi,\phi)} = S^{n_a} ,    \hspace{5mm} 
\sup_{\phi \in {\cal H} \otimes {\cal H} }\frac{| ( \phi ,S_X^{i(b),1} S_X^{i(b),2} \phi)|}{(\phi,\phi)} = S^{2n_b}.
$$  
%We consider a term of random Hamiltonian as a function of a sequence $O^a=(O^{a}_X)_{ X \in C_L^a }$ of spin operators and 
%the random variables $(g_X^a)_{X \in C_L^a}$ with an arbitrarily fixed index $a \in A$.\\

\noindent
{\bf Note} { \it The density $m^0_L$ is identical to the overlap
\begin{equation} 
m^0_L = R_{1,2}^c, \label{note0}
\end{equation} 
and the replica symmetry at $(J_0^0,J_1^0)=(0,0)$ dose not imply $\langle m^0_L\rangle =0$, 
but implies $\langle R_{1,2}^c \rangle =
\langle R_{k,l} ^c\rangle$ for $1 \leq k < l \leq n $.}\\

Here we employ the definition of replica symmetry breaking for disordered Ising systems given by Chatterjee \cite{C2}
also for quantum systems.
For an arbitrary sequence of  $(g_X^a)%_{X\in C_L^a,a\in A}
$of random variables,
we say that 
a replica symmetry breaking  occurs, if %the overlap operator has finite variance
 \begin{eqnarray}
 \lim_{L \rightarrow \infty }  
\langle (R_{1,2}^a -  {\mathbb E} \langle R_{1,2}^a \rangle)^2 \rangle \neq 0,
\label{RSB2}
\end{eqnarray}
for $ \exists a \in A$ and for $(J_0^0,J_1^0)=(0,0).$
%} This definition of replica symmetry breaking is

Next,  we remark the nature of spontaneous replica symmetry breaking.
 %{\definition  
For an arbitrary  sequence $(g_X^a)%_{X\in C_L^a,a\in A}
$ of random variables 
we can say that a spontaneous replica symmetry breaking occurs, if %the following limit operations do note commute
\begin{eqnarray} 
 \lim_{J_0^0\rightarrow 0}   \lim_{J_1^0\rightarrow 0} \lim_{L \rightarrow \infty}\langle{R_{1,2}^a}^k \rangle 
 \neq \lim_{L \rightarrow \infty} 
 \lim_{J_0^0 \rightarrow 0} \lim_{J_1^0\rightarrow 0} \langle {R_{1,2}^a} ^k\rangle,
\label{SRSB}
\end{eqnarray}
for $\exists r \in \{0,1\},$ 
for $ \exists a \in A$ and for $ \exists k >0$. 
%}\\

%We show two theorems for spontaneous symmetry breaking in the following.\\
Now, we apply Theorem \ref{MT3} and \ref{MT4} to $m_L^0=R_{1,2}^c$.  
If (I) in Theorem \ref{MT3} occurs in some samples, namely 
$$
\lim_{L \rightarrow \infty }%\lim_{J_0^0 \rightarrow 0} \lim_{J_1^0 \rightarrow 0}
 \langle (R_{1,2}^c-
\langle R_{1,2}^c\rangle  )^2 \rangle\neq 0,
$$
at $(J_0^0,J_1^0)=(0,0)$, then (II) the limit operations are non-commutative and
the spontaneous replica symmetry breaking occurs in such samples.
This result corresponds to the Koma-Tasaki theorem which shows the 
detection of the spontaneous  symmetry breaking by the long range order in symmetric Gibbs state
in deterministic quantum systems. 

Theorem \ref{MT4} for $m_L^0=R_{1,2}^c$ implies that 
the replica symmetry breaking 
%implies  that spontaneous replica symmetry breaking is not rare. Therefore, the replica symmetry breaking 
should be spontaneous replica symmetry breaking.  
If (III) in Theorem \ref{MT4} occurs in some samples, namely 
$$
\lim_{L \rightarrow \infty }%\lim_{J_0^0 \rightarrow 0} \lim_{J_1^0 \rightarrow 0}
 \langle (R_{1,2}^c-{\mathbb E}
\langle R_{1,2}^c\rangle  )^2 \rangle\neq 0,
$$
in the replica symmetric calculation, % for all $(g_X^a) \in {\cal T}_K$, 
then (IV) and the spontaneous replica symmetry breaking occurs in such samples for almost everywhere in the coupling constant space.  
In each sample, one of the  replica symmetry breaking Gibbs states realizes, 
and the overlap operator becomes self-averaging. 
Theorem \ref{MT4} implies also that if the spontaneous replica symmetry breaking is a rare event,  
then the replica symmetry breaking rarely occurs, namely
$R_{1,2}^c$ is self-averaging even in the replica symmetric expectation. 
\\

Acknowledgment

It is pleasure to thank I. Affleck, T. Koma and R. M. Woloshyn  for helpful discussions in early stage of this study.
I would like to thank I. Affleck for kind hospitality at UBC. 
Also, I am grateful to R. M. Woloshyn for careful reading of the manuscript and for kind hospitality at TRIUMF.
%when the spontaneous replica symmetry breaking occurs. 
%(I) in Theorem \ref{MT4} implies that a spontaneous replica symmetry breaking is rare event in all disordered samples.
%Even though the replica symmetry breaking yields the finite variance of the overlap in
%the replica symmetric expectation, this variance vanishes for any inter replica coupling $J^0_1 \neq 0$ defined by (\ref{inthamil}).    
%\newpage

\end{document}